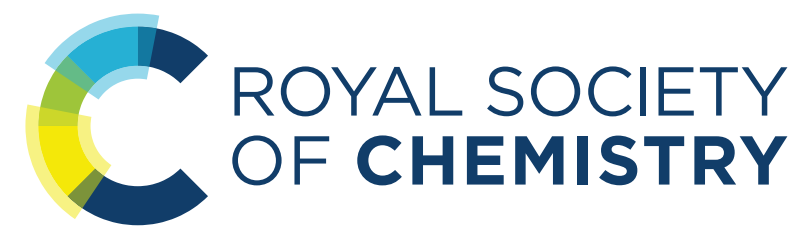

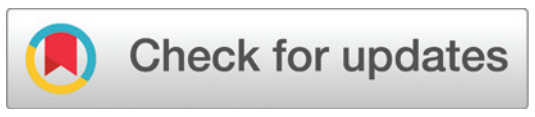

# Resonance Raman scattering and anomalous anti-Stokes phenomena in CrSBr†

Satyam Sahu, *[a,b] Charlotte Berrezueta-Palacios,[c] Sabrina Juergensen,[c] Kseniia Mosina,[d] Zdeněk Sofer, [d] Matěj Velický, *[a] Patryk Kusch *[c] and Otakar Frank *[a]

CrSBr, a van der Waals material, stands out as an air-stable magnetic semiconductor with appealing intrinsic properties such as crystalline anisotropy, quasi-1D electronic characteristics, layer-dependent antiferromagnetism, and non-linear optical effects. We investigate the differences between the absorption and emission spectra, focusing on the origin of the emission peak near 1.7 eV observed in the photoluminescence spectrum of CrSBr. Our findings are corroborated by excitation energy-dependent Raman experiments. Additionally, we explore the anti-Stokes Raman spectra and observe an anomalously high anti-Stokes-to-Stokes intensity ratio of up to 0.8, which varies significantly with excitation laser power and crystallographic orientation relative to the polarization of the scattered light, highlighting the unique vibrational and electronic interactions in CrSBr. Lastly, we examine stimulated Raman scattering and calculate the Raman gain in CrSBr, which attains a value of $1 \times 10^8$ cm GW$^{-1}$, nearly four orders of magnitude higher than that of previously studied three-dimensional systems.

## 1 Introduction

Since the discovery of magnetism in $CrI_3$[1] and $Cr_2Ge_2Te_6$,[2] van der Waals magnets have garnered significant attention for both fundamental research and their potential applications in modern semiconductor technology and spintronic devices.[3–9] However, many of these materials pose challenges associated with degradation under ambient conditions, which hinder their fabrication and utilization.[2,10–12] Recently, the discovery of CrSBr[13,14] marked a breakthrough in the search for air-stable van der Waals magnets, offering a promising alternative.

CrSBr crystallizes in an orthorhombic structure, where each Cr atom is surrounded by four chalcogen and two halogen atoms, forming a distorted octahedral coordination.[15,16] This material is a two-dimensional antiferromagnetic semiconductor with a direct electronic band gap at the Γ point in the Brillouin zone, with a broadly varied energy range of 1.5–2.3 eV, reported in the literature, while it is expected to have an optical gap of ~1.3 eV.[15,17–20] It hosts tightly bound magneto-excitons,[15,17,21–25] and its unique combination of semiconducting and magnetic properties, along with its quasi-1D electronic nature,[18] positions it as an ideal candidate for magneto-optical,[26] magneto-transport,[27] and polarization-sensitive[28] applications. Furthermore, significant magnetoresistance has been observed below its magnetic ordering temperature (Néel temperature).[17] An additional intriguing feature of CrSBr is its unusual anisotropic thermal expansion, with a negative thermal expansion coefficient along the a-axis and a positive one along the b-axis.[29]

Despite extensive research on CrSBr, there is no consensus on the size of its optical and electronic band gaps, where experimental results and theoretical calculations often yield conflicting values. These discrepancies may arise from differences in the measurement techniques, sample quality, or theoretical models. Moreover, the absorption spectrum of CrSBr[30] is markedly shifted relative to its emission spectrum,[18,28] a phenomenon indicative of complex electronic transitions or defect states that are yet to be thoroughly investigated. The unexplored anti-Stokes Raman region and non-linear Raman effects could also offer further insights into its phonon dynamics and anharmonicity.

This work addresses the absorption–emission disparity through detailed optical characterization, including absorp-tion, photoluminescence (PL), and resonance Raman spectroscopy (RRS). To strengthen our conclusions, we employ excitation energy-dependent Raman profile measurements, which link the optical responses to specific vibrational modes. The combined results suggest that the higher energy PL emission is indirect in nature and, therefore, occurs at lower transition energies compared to optical absorption and RRS. Our study also reveals a unique anti-Stokes Raman behavior of CrSBr, with the anti-Stokes-to-Stokes intensity ratio ($I_{AS}/I_S$) reaching values of up to 0.8 at room temperature, and reports the first observation of stimulated Raman scattering (SRS) in CrSBr, accompanied by a remarkably high Raman gain factor of $10^8$ cm $GW^{-1}$. These results not only expand our understanding of CrSBr but also underscore its potential for applications such as Raman lasers.

[a]J. Heyrovský Institute of Physical Chemistry, Czech Academy of Sciences, Dolejškova 2155/3, 182 23 Prague, Czech Republic. E-mail: satyam.sahu@jh-inst.cas.cz, matej.velicky@jh-inst.cas.cz, otakar.frank@jh-inst.cas.cz

[b]Department of Biophysics, Chemical and Macromolecular Physics, Faculty of Mathematics and Physics, Charles University, Ke Karlovu 3, 121 16 Prague, Czech Republic

[c]Institut für Physik, Freie Universität Berlin, Arnimallee 14, 14195 Berlin, Germany. E-mail: patryk.kusch@fu-berlin.de

[d]Department of Inorganic Chemistry, University of Chemistry and Technology Prague, Technická 5, 166 28 Prague, Czech Republic

† Electronic supplementary information (ESI) available. See DOI: **https://doi.org/10.1039/d5nr00562k**

# 2 Experimental

CrSBr bulk crystals were grown by chemical vapor transport method as reported elsewhere.[25] Bulk flakes were exfoliated using the traditional mechanical exfoliation method onto polydimethylsiloxane and were subsequently transferred over periodic arrays of 5 μm diameter holes in a 200 nm thick silicon nitride membrane (TED PELLA, INC. Prod No. 21536-10) by a standard dry-stamp transfer method as reported in our previous work.[28]

The Raman (PL) measurements were carried out on WITEC Alpha 300R (Witec Instruments, Germany) with 1800 lines per mm (600 lines per mm) grating. The 1.96 eV and 2.33 eV excitation laser beams were focused through a 100× objective (NA = 0.9) lens in a backscattering geometry. We used a low-frequency filter set to obtain both Stokes and anti-Stokes Raman components. The incident laser power was controlled using a variable neutral density filter.

The absorption spectra were measured at room temperature using a custom-built system. This setup featured a broad-band light source (NKT – FIU 15), with the light directed into an inverted Olympus microscope. A 100× objective lens (NA = 0.9) focused the light onto the sample and delivered the reflected light through an optical fiber to an Avantes spectrometer.

For resonant Raman experiments, we used a continuous wave dye ring laser (Radiant dyes Laser GmbH; available excitation range: 550 nm–690 nm) and a tunable Coherent Ti : Sa ring laser (690–1050 nm). Raman spectra were recorded with a Horiba T64000 triple monochromator (equipped with 900 lines per mm) in backscattering configuration, coupled with a microscope, motorized XY stage, and silicon CCD. Resonant micro-Raman spectroscopy of CrSBr was performed at room temperature using a 100× objective (NA = 0.8) for excitation wavelengths from 1.65 eV to 2.34 eV with light polarized parallel to the crystal axes, which were identified using the polarization dependent Raman spectroscopy of CrSBr flake under 2.33 eV excitation energy.[28] All the RRS measurements were conducted on the same sample spot and with the same laser power of 1 mW for all excitation wavelengths. Raman intensity calibration was performed using the silicon Raman mode at 523 $cm^{-1}$, ensuring a consistent Raman cross-section across the wavelength range.

# 3 Results and discussion

## 3.1 Characterization

The optical microscope image of a suspended 55 nm-thick CrSBr flake transferred onto 5 μm-diameter holes etched in a $Si_3N_4$ substrate is presented in Fig. 1a. To investigate the optical and electronic properties of CrSBr, we analyzed its optical transitions using PL and absorption measurements (Fig. 1b and c). Fig. 1b displays the PL spectrum of the flake shown in Fig. 1a under 2.33 eV excitation for the laser polarization along the a- and b-crystalline axes at room temperature. As reported earlier, the difference in the peak intensities arises due to the anisotropic nature of CrSBr.[31,32] The PL spectra feature two distinct emission peaks, positioned at around 1.31 eV and 1.72 eV. The low-energy peak (1.31 eV) is linked with the direct band-to-band excitonic transition at the Γ point in the Brillouin zone.[31] The higher energy emission (1.72 eV) has been associated with the transition from the conduction band minimum to the lower component of the spin-split valence band ($\Delta_{s\text{-}o}$ = 0.4 eV) at the Γ point in the literature.[32] The absorption spectrum of CrSBr is shown in Fig. 1c. In the spectrum, we observe an asymmetric absorption peak at 1.85 eV, with a shoulder at 2.26 eV. In contrast to the PL spectrum, there is no peak in the vicinity of 1.72 eV. We could not access the low-energy range (below ∼1.35 eV) due to the decreased sensitivity of our setup.

In the PL and absorption measurements, we made two key observations: first, the separation between the PL peaks and the absorption peaks is 0.41 eV, which is in line with the reported valence band splitting energy ($\Delta_{s\text{-}o}$ = 0.4 eV (ref. 32)). Therefore, it is highly probable that the valence bands at the Γ point play a role in both the absorption and emission processes. Second, the absence of absorption at ∼1.7 eV indicates that this emission is not due to the direct band-to-band recombination at the Γ point but is probably due to some phonon-assisted indirect transitions, also suggested in the recent work by Shen *et al.*[33]

Moving forward, we measured the Raman spectra of CrSBr flakes using two (2.33 eV and 1.96 eV) excitation energies, $E_L$. Fig. 1d shows the representative Raman spectra that contain three Stokes peaks, namely $A_g^1$, $A_g^2$, and $A_g^3$ modes as reported in the previous study.[34] When exciting the sample with $E_L$ = 2.33 eV polarized along the crystal's a-axis, we observed only a single mode ($A_g^2$) with a clear Lorentzian lineshape at 244.7 $cm^{-1}$, whereas, for the polarization along the b-axis, all three modes with frequencies of 113.6 $cm^{-1}$, 244.1 $cm^{-1}$, and 342.6 $cm^{-1}$ were visible due to CrSBr's crystalline anisotropy.[18,34] On the contrary, for $E_L$ = 1.96 eV, we observe a slight asymmetry in the Raman modes, especially for the $A_g^2$ mode (along the b-axis), whose appearance resembles a Fano lineshape (ESI Fig. S1†) that has been earlier attributed to a van Hove singularity from the quasi-1D electronic nature of CrSBr.[18]

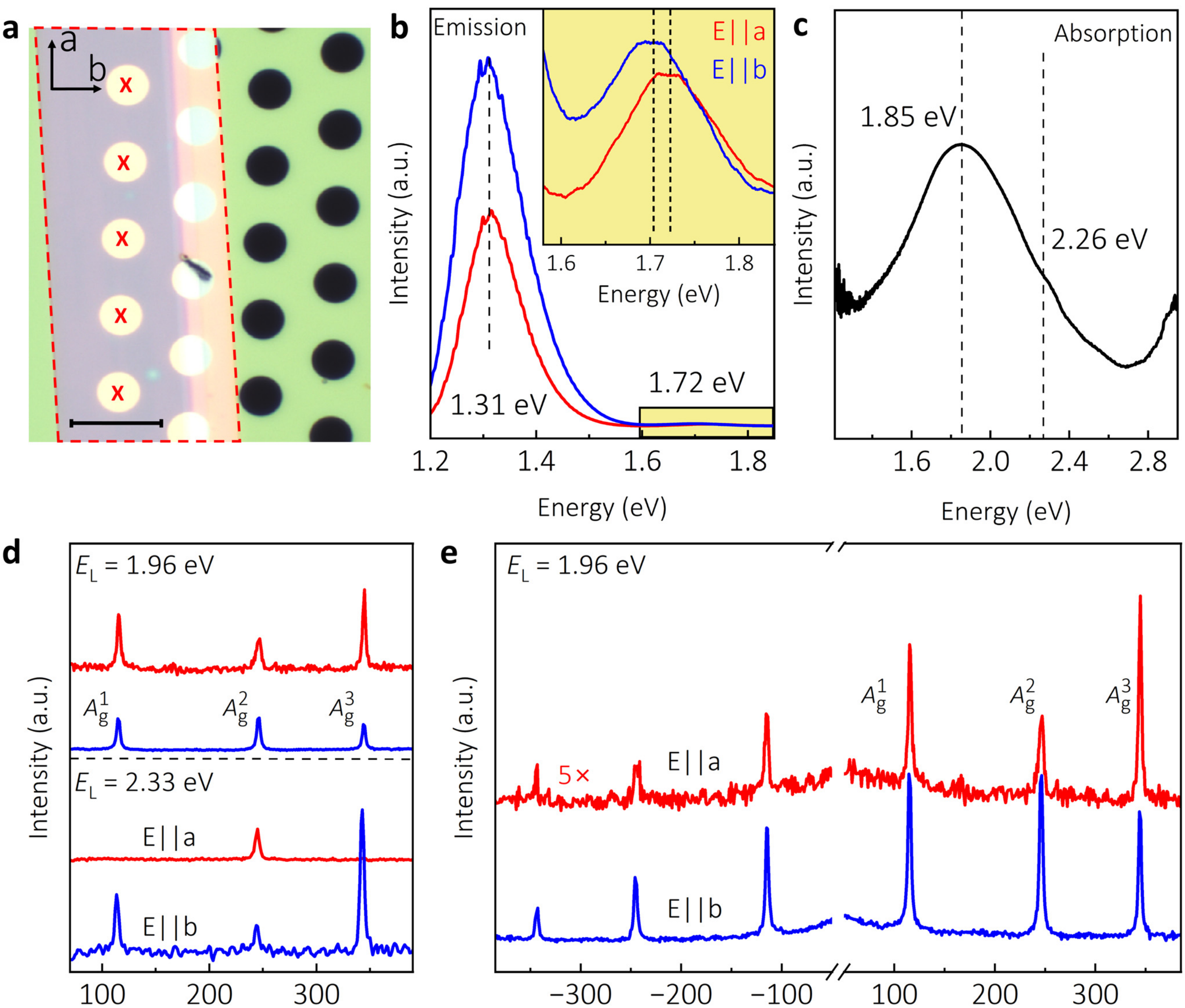

**Fig. 1** Optical spectroscopy and characterization: (a) Optical image of 55 nm-thick suspended CrSBr flake on a $Si_3N_4$ grid with 5 μm-diameter holes. The black arrows indicate the crystalline a- and b-axes orientations, scale bar = 10 μm. Holes marked with X were used for the measurements. (b) Photoluminescence spectrum of CrSBr for the 2.33 eV laser polarized along a- and b-axes, respectively. Inset: magnified view of the highlighted emission at 1.72 eV. (c) Absorption spectrum of CrSBr. Dashed lines show the two optical transitions at 1.85 eV and 2.26 eV. (d) Raman spectra of CrSBr under 1.96 eV and 2.33 eV excitation energy at 100 μW. All three modes were observed in the case of the 1.96 eV excitation (for both orientations) and the 2.33 eV excitation (along the b-axis), while only $A_g^2$ mode is visible for the 2.33 eV excitation along the a-axis. (e) Raman spectra under 1.96 eV at 100 μW excitation along the a- and b-axes, respectively, showing all three Raman modes both in Stokes and anti-Stokes regions.

While all three modes are due to out-of-plane atomic displacements, the lowest frequency $A_g^1$ mode in CrSBr has a significant contribution of Br atoms to the phonon density of states since it resides at the top and bottom of each layer, with sulfur and chromium atoms in the middle. Moreover, due to the position of Br atoms within the lattice, the $A_g^1$ mode mediates the interlayer coupling between the layers and is more sensitive to external factors.[14] The $A_g^2$ and $A_g^3$ modes have a larger contribution of Cr and S atoms, respectively, to the phonon density of states, and hence these modes mediate the intralayer coupling.[14]

In addition to enabling the observation of low-intensity modes and asymmetry in Raman peaks, we have found that the excitation energy has dramatic effects on the Stokes and anti-Stokes Raman signatures. Fig. 1e shows both the Stokes and anti-Stokes Raman of CrSBr under 1.96 eV excitation along the a- and b-axes. We observe that the anti-Stokes peaks are quite strong, an unusual occurrence not present in most materials,[35,36] except in the case of $WSe_2/WS_2$ vertical heterostructure, where the anti-Stokes signal was reasonably strong.[37] The anomalously intense anti-Stokes peaks suggest the possibility of additional phenomena taking place in the Raman process, namely the resonance effect, electron–phonon, and exciton–phonon coupling in CrSBr.

### 3.2 Resonance Raman spectroscopy

To shed light on the observed discrepancy in the PL and absorption spectra, we employed RRS measurements. By selecting appropriate excitation energies and polarization conditions, RRS can provide detailed information on various electronic transitions, symmetries, and selection rules.[38–40] We recorded the Stokes and anti-Stokes Raman spectra as a function of the excitation energy in the range from 1.7 eV to 2.45 eV. In the Stokes RRS process, incoming resonance enhances the signal when the incident photon energy matches a direct band-to-band transition. Outgoing resonance occurs when an incident photon excites an electron to an intermediate state, followed by inelastic scattering and recombination, producing a Raman-shifted photon. In RRS, for longitudinal and transverse optical phonons (including the $A_g^1$, $A_g^2$, and $A_g^3$ modes in our study), the dominant processes involve direct excitations and electron recombination at zero momentum.[40] The anti-Stokes process involves the annihilation of an existing phonon, with an outgoing resonance appearing one phonon energy below the band-to-band transition energy.

In Fig. 2, we present the RRS data of all three modes for both Stokes and anti-Stokes spectra, where we plot the integrated intensity as a function of the excitation energy. For all the modes, we observe two resonances at around 2.0 eV and 2.2 eV (Table 1). For both Stokes and anti-Stokes components, we observe the symmetric shape of the individual fitted resonance peaks due to the identical intensity of the incoming and outgoing resonance.[41,42] The phonon energy is small (~14 meV for $A_g^1$ mode), much smaller than the width of the whole resonance profile (~0.3 eV), and even smaller than the individual resonances.[43] Therefore, the incoming and outgoing resonances are not resolved in either of the peaks. The resonance energies are obtained by fitting the scattering intensity:[43]

$$I \sim \left| \frac{M_1}{E_L - E_R - i\frac{\gamma}{2}} - \frac{M_2}{E_L - E_R - \hbar\omega_{ph} - i\frac{\gamma}{2}} \right|^2 \quad (1)$$

where $E_L$ is the excitation, and $E_R$ is the resonance energy. $M_1$ and $M_2$ are the matrix elements, $\hbar\omega_{ph}$ is the phonon energy, and $\frac{\gamma}{2}$ represents the finite lifetime broadening. The first term in eqn (1) is responsible for the incoming resonance, and the second term for the outgoing resonance. To fit a symmetric cross-section, we set the combined matrix elements $M_1 = M_2$. Separate fits of the Stokes and anti-Stokes resonance profiles yielded similar resonance energies for the same lifetime broadening (see Table 1) for Stokes and anti-Stokes components of the respective phonon modes. The resonance energies of the anti-Stokes modes are lower than those of the Stokes ones, with the separations between them approximately corresponding to the respective phonon energies when the fitting errors are taken into account. Such difference is expected when comparing resonances between anti-Stokes and Stokes modes.[44]

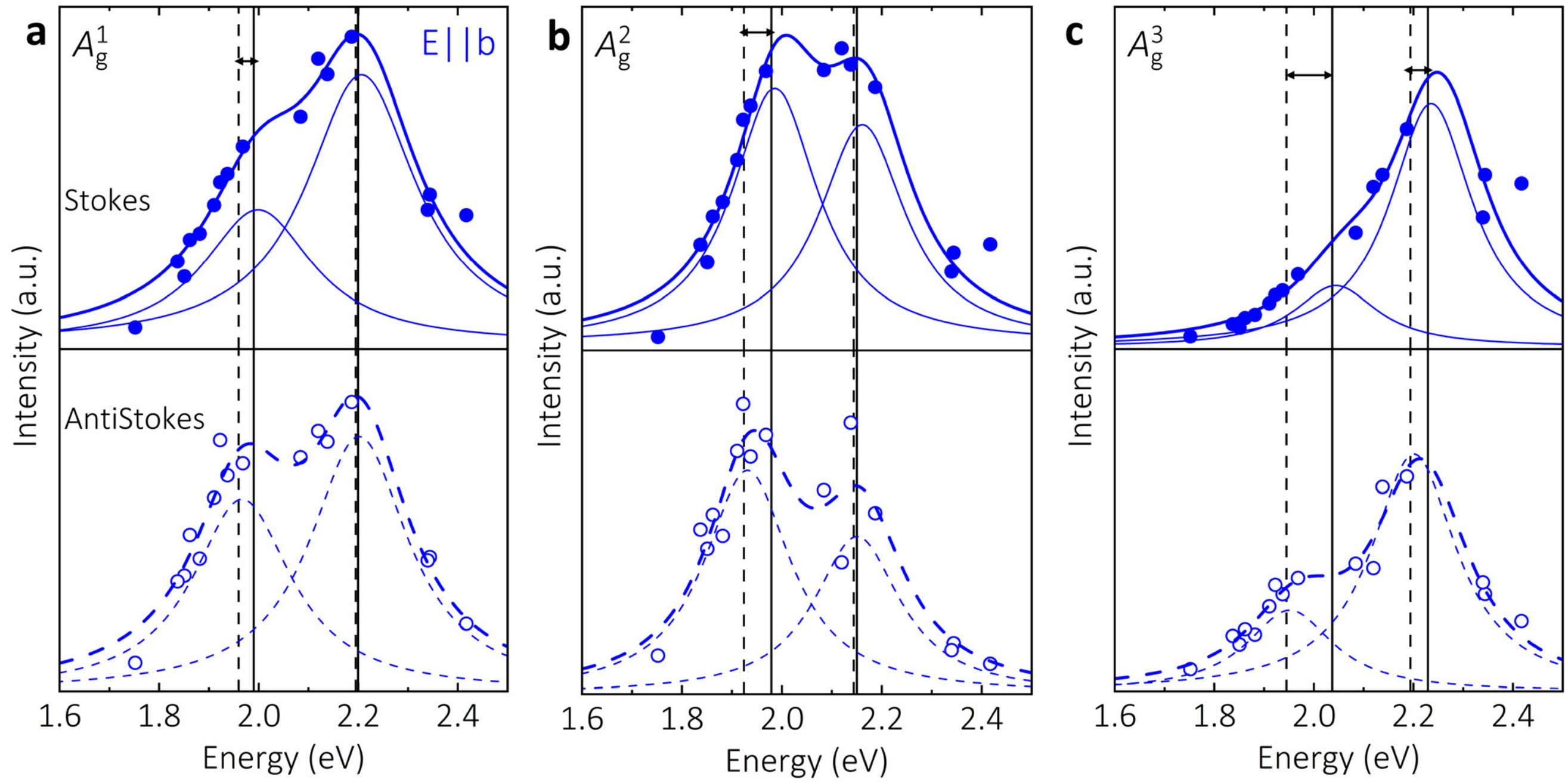

Fig. 2 Resonant Raman excitation profile: Resonant Raman profiles of the (a) $A_g^1$, (b) $A_g^2$, and (c) $A_g^3$ modes for both Stokes (solid circles and curves) and anti-Stokes (empty circles and dashed curves) regions for laser co-polarized along b-axis. The symbols denote the experimental data points, while the solid and dashed blue curves are the fits using eqn (1).

**Table 1** Resonance positions ($E_R$) extracted from the fits

| | Raman shift (in eV) [in cm$^{-1}$] | $E_{R1}$ | | $E_{R2}$ | | $\gamma/2$ | |
|---|---|---|---|---|---|---|---|
| | | Stokes (eV) | anti-Stokes (eV) | Stokes (eV) | anti-Stokes (eV) | Stokes (eV) | anti-Stokes (eV) |
| $A_g^1$ | 0.014 [113.6] | 1.99 ± 0.02 | 1.96 ± 0.01 | 2.20 ± 0.01 | 2.20 ± 0.01 | 0.13 ± 0.00 | 0.12 ± 0.01 |
| $A_g^2$ | 0.030 [244.1] | 1.98 ± 0.02 | 1.92 ± 0.02 | 2.15 ± 0.02 | 2.14 ± 0.03 | 0.10 ± 0.01 | 0.10 ± 0.00 |
| $A_g^3$ | 0.043 [342.6] | 2.04 ± 0.05 | 1.94 ± 0.02 | 2.23 ± 0.01 | 2.19 ± 0.01 | 0.10 ± 0.00 | 0.10 ± 0.01 |

The resonances observed at approximately 2.0 eV and 2.2 eV align well with the absorption measurements, indicating the presence of direct electronic transitions in these energy ranges. In contrast, no resonance is detected near 1.7 eV, implying the absence of a direct transition at that energy. Consequently, the PL at 1.72 eV cannot be attributed to a direct transition between the spin-split valence band and the conduction band minimum at the Γ point. Instead, it is likely the result of an indirect transition supporting our hypothesis and the recent work by Shen *et al.*[33]

We explain the slight discrepancy between the Raman excitation profiles and absorption spectra by noting that, in RRS, selection rules primarily favor phonons at the Γ point. As a result, the electronic transitions occur predominantly at the Γ point, whereas absorption integrates contributions from the entire Brillouin zone. Finally, for energies < 1.3 eV, the sensitivity of our setup and the Raman cross-section of CrSBr are too low to detect Raman signals effectively.

### 3.3 Anti-Stokes Raman spectra

Next, we shift our focus to the anomalous anti-Stokes Raman modes. We performed laser power-dependent Raman measurements and fitted each spectrum with a Lorentzian line shape on top of a linear background. The average between the measured Stokes and anti-Stokes peak positions of each Raman mode is used to determine the corresponding peak positions and to correct for the zero-point offset of the spectrometer. As the laser power increases, the integrated areal intensity of the anti-Stokes mode with respect to the Stokes mode increases due to an increase in the phonon population. This is accompanied by a softening of all three modes, as shown in Fig. 3(a–c). Similar frequency red shifts have been observed in power- and temperature-dependent Raman measurements of a wide variety of materials[45–47] and can be attributed to the anharmonic (phonon–phonon interactions) terms of the lattice potential energy.[48]

Upon careful examination of the Raman peak position *versus* laser power graphs for all modes, an interesting pattern can be noticed: when the laser (and scattering) polarization aligns with the crystalline a-axis, the slopes of the vibrational frequencies are markedly less negative compared to the laser polarization along the b-axis. This behavior can be rationalized by the opposite thermal expansion coefficients along the a- and b-axes.[29] The negative ($-6.4 \times 10^{-6}$ K$^{-1}$)[29] thermal expansion along the a-axis and the positive ($10.9 \times 10^{-6}$ K$^{-1}$)[29] expansion along the b-axis contribute to an increase and decrease, respectively, in the Raman mode frequency with rising temperature or laser power. However, this effect is small compared to the frequency downshifts caused by the increasing anharmonic interactions with a rise in temperature, leading to overall negative slopes, which only vary in magnitude for the two orientations.

The anti-Stokes-to-Stokes ratio contains information on the phonon populations and their temperatures. We extract the anti-Stokes and Stokes peak intensities for all three modes from the fits and plot the ratio $I_{AS}/I_S$ as a function of excitation power at 295 K in Fig. 3(a–c).

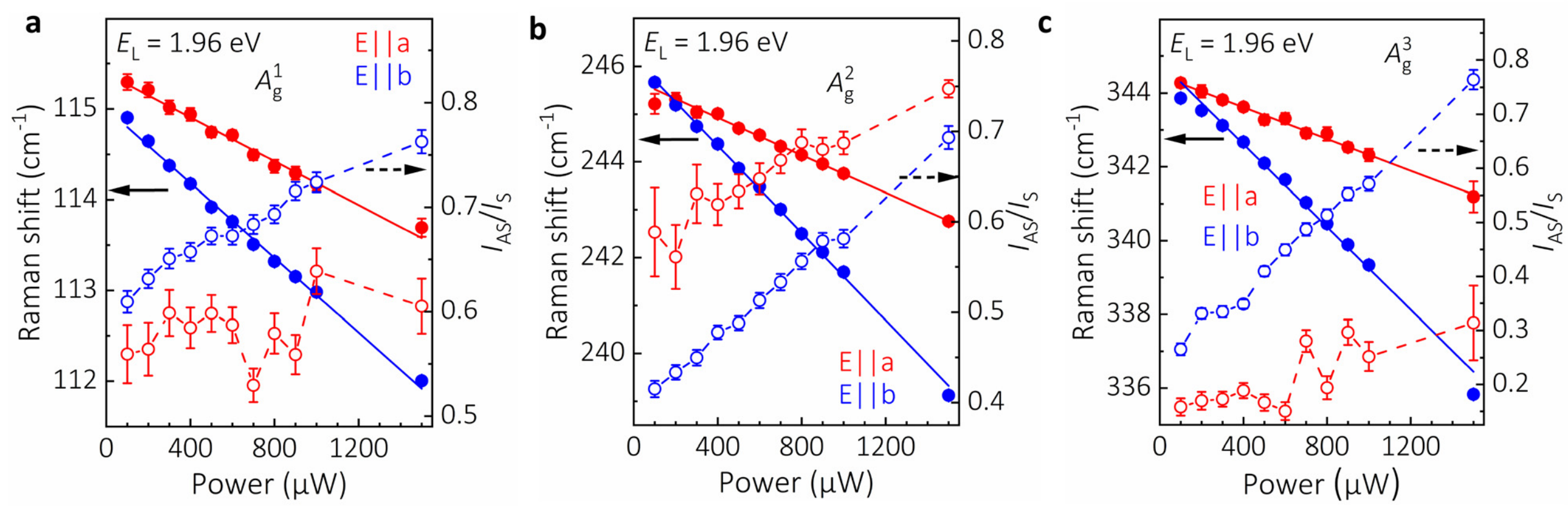

**Fig. 3** Anomalous anti-Stokes to Stokes ratio: Power dependence of peak shifts (solid circles) and anti-Stokes-to-Stokes intensity ratio (empty circles) along a- and b-axes for (a) $A_g^1$, (b) $A_g^2$, and (c) $A_g^3$ modes under $E_L$ = 1.96 eV excitation energy. The solid lines are linear fits, while the dashed ones are data point connectors. Error bars correspond to the fitting error for the Raman shift and to the variance for the anti-Stokes-to-Stokes ratio.

Here, we also observe two interesting features. First, the $I_{AS}/I_S$ for $A_g^1$ and $A_g^3$ is higher for the laser polarization along the b-axis, corresponding to the strong polarization of these two modes along this direction (*cf.*, Fig. 1). Conversely, for the $A_g^2$, the $I_{AS}/I_S$ is larger for the polarization along the a-axis; again, in line with its strong polarization dependence. Second, the $I_{AS}/I_S$ for $A_g^3$ mode abruptly increases from ~0.3 to ~0.8 (at 1500 μW) on aligning the laser polarization from the a- to the b-axis, whereas for $A_g^1$ and $A_g^2$, the $I_{AS}/I_S$ changes only by ≈0.15 and ≈0.05, respectively. Such polarization dependence of $I_{AS}/I_S$ can again be attributed to the crystallographic anisotropy of CrSBr and the different resonance conditions for the individual modes, *cf.*, the relative intensities of the two resonance peaks in Fig. 2.

The anomalous anti-Stokes-to-Stokes ratio observed under 1.96 eV excitation can be attributed to a resonance effect previously identified near 2.0 eV and 2.2 eV across all three Raman modes. Shifting the excitation energy away from the resonance region results in a decrease in the anti-Stokes-to-Stokes ratio (ESI Fig. S2†), further supporting the role of the strong resonance effect in this phenomenon. Additionally, the distinctive 1D crystal structure of CrSBr enhances this effect through pronounced exciton–phonon[31] and electron–phonon[49] coupling, as reported in the literature.

### 3.4 Stimulated Raman scattering

In the previous section, we observed a distinctive trend in the $A_g^2$ mode intensity for the laser polarized along a-axis as the laser power increased (see Fig. 4 and ESI Fig. S3†). This trend can be understood through the principles of SRS, a nonlinear optical process where the intensity of the Raman signal depends on the laser power ($P$). In general, the intensity of the Raman signal follows a power law, given by $I \propto P^n$, where $n$ is the exponent that characterizes the relationship between the intensity and laser power. For normal (spontaneous) Raman scattering, $n$ = 1, meaning the intensity increases linearly with the laser power. However, in the case of SRS, $n > 1$, indicating a non-linear dependence of the Raman intensity on the laser power.[50]

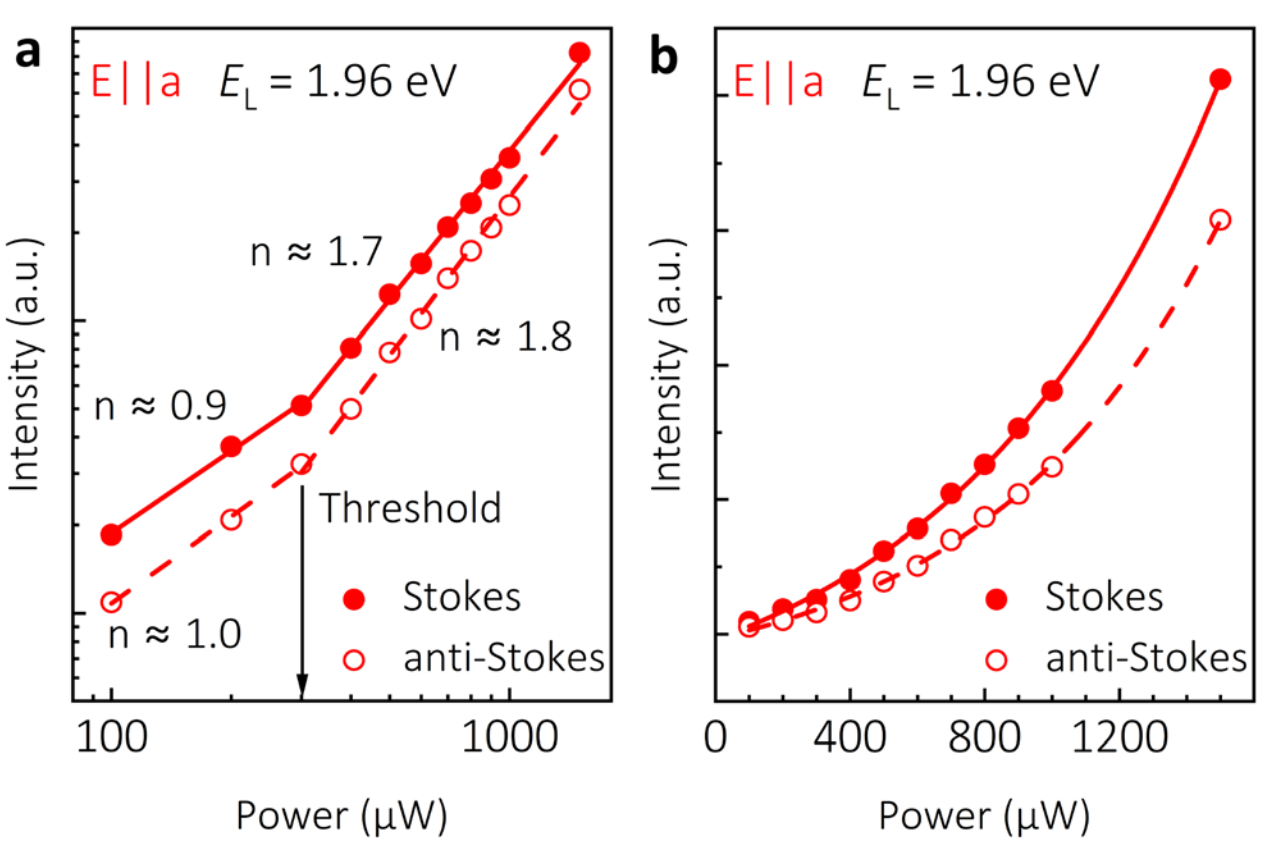

Fig. 4 Stimulated Raman scattering: Integrated Stokes and anti-Stokes $A_g^2$ intensities *vs.* laser power under $E_L$ = 1.96 eV excitation in (a) log–log and (b) linear–linear scales, respectively. Symbols denote the data points, while the solid and dashed lines are the linear (a) and exponential (b) fits, respectively. The laser power threshold for the SRS process was 300 μW.

As shown in Fig. 4a, CrSBr exhibits spontaneous Raman scattering with $n$ = 0.93 ± 0.05 for the Stokes and 0.98 ± 0.04 for the anti-Stokes components in the low-power range up to 300 μW. However, above this power threshold, both the Stokes and anti-Stokes intensities were observed to exhibit strongly nonlinear behavior with $n$ = 1.69 ± 0.03 and $n$ = 1.80 ± 0.04, respectively, consistent with the manifestation of SRS. To rule out potential structural changes to the CrSBr by laser-induced heating, the Raman intensities were also monitored by decreasing the power back to 100 μW. No hysteresis was observed in the experiment (ESI Fig. S4†).

Using the exponential law, $I \propto e^{g \cdot L \cdot I_{\text{in}}}$, for the Raman intensity, we determined the SRS gain, $g$, to be ≈1 × $10^8$ cm $GW^{-1}$ for both Stokes and anti-Stokes modes from the fits in Fig. 4b. Here, $I_{\text{in}}$ is the power density of the incident laser light (= $P$/laser spot area) and $L$ (= 1 μm) is the interaction length of the Raman medium.[51] The obtained gain is twice as high as compared to the recently reported Si nanowires,[52] four orders higher than that reported for GaInAs/AlInAs quantum cascade Raman lasers, and more than seven orders of magnitude higher than that reported for bulk materials such as Si Raman lasers and other Raman crystals but an order of magnitude smaller than that of single-walled nanotubes suggesting that the nanotubes exhibit even stronger nonlinear effects.[51,53–55]

Such strong nonlinear effects in CrSBr can be explained in the context of its quasi-1D rod-like structure, where the movement of electrons is strictly confined along only one direction. Thus, all of the light-generated electrons are expected to be involved in the electron–phonon interaction and result in significant amplification of phonons.[51,56] The threshold laser power to achieve SRS in CrSBr was comparable to that reported in Si nanowires (~30 kW $cm^{-2}$),[52] which is several orders of magnitude lower in comparison to the bulk Si-based Raman lasers.[55] In addition, the finite size of the crystal can lead to cavity-like enhancement effects, as reported for GaP or Ge nanowires.[50,57]

Interestingly, while we do not observe any SRS for the $A_g^2$ mode for laser polarization along the b-axis, it does exhibit Fano asymmetry (Fig. 1). The observation of SRS for laser polarization along the a-axis and Fano resonance along the b-axis can be attributed to the symmetry and orientation of the Raman tensor elements that mediate the interaction of the laser electric field vector with the material.[58,59] When the laser is polarized along the a-axis, the electric field aligns more effectively with Raman tensor elements that enhance the SRS process, facilitating coherent amplification of specific vibrational modes.[58] Conversely, along the b-axis, the coupling between the laser field and the Raman tensor elements interacts with electronic resonances or continuum states, producing interference effects characteristic of Fano resonance.[59]

For the remaining two modes, $A_g^1$ and $A_g^3$, we observe neither SRS nor Fano asymmetry in any of the polarization directions, which confirms the peculiar electron–phonon interactions specific for the $A_g^2$ mode.[18,34]

## 4 Conclusions

In conclusion, our study provides insights into the disparity between absorption and emission energies in CrSBr, attributing the emission at 1.72 eV to an indirect optical transition and the absorptions at 1.85 and 2.26 eV to direct transitions as supported by resonance Raman spectroscopy. The strong anti-Stokes-to-Stokes intensity ratios and the observed stimulated Raman scattering highlight the influence of resonance effects and electron–phonon coupling. These findings underscore CrSBr's potential in nonlinear optical applications. By enhancing our understanding of its nonlinear behavior, this work lays a solid foundation for further exploration of CrSBr in advanced photonic technologies.

## Data availability

The data and analyses underlying this study are available from the HeyRACK repository at **https://doi.org/10.48700/datst.e94rx-q2q41**.

## Conflicts of interest

There are no conflicts to declare.

## Acknowledgements

O. F., C. B. P., and P. K. acknowledge the support of the Czech Science Foundation Project No. 23-06174K. P. K. also acknowledges the DFG for funding (KU4034 2-1). Z. S. and K. M. were supported by the ERC CZ program (project LL2101) from Ministry of Education Youth and Sports. M. V. acknowledges the support of the Lumina Quaeruntur fellowship No. LQ200402201 by the Czech Academy of Sciences. S. S. thanks the grant No. SVV-2023-260716 from Charles University. This work was also supported by the SupraFAB Research Facility and the Focus Area NanoScale at Freie Universität Berlin, and by the Ministry of Education, Youth, and Sports of the Czech Republic, Project No. CZ.02.01.01/00/22_008/0004558, co-funded by the European Union.